# Session-based Recommendation with Self-Attention Networks

Jun Fang[1]

1. School of Communication and Information Engineering, Shanghai University, Shanghai 200444, China;

fj6114@gmail.com

**Abstract:** Session-based recommendation aims to predict user's next behavior from current session and previous anonymous sessions. Capturing long-range dependencies between items is a vital challenge in session-based recommendation. A novel approach is proposed for session-based recommendation with self-attention networks (SR-SAN) as a remedy. The self-attention networks (SAN) allow SR-SAN capture the global dependencies among all items of a session regardless of their distance. In SR-SAN, a single item latent vector is used to capture both current interest and global interest instead of session embedding which is composed of current interest embedding and global interest embedding. Some experiments have been performed on some open benchmark datasets. Experimental results show that the proposed method outperforms some state-of-the-arts by comparisons.

## 1. Introduction

With the ever-growing volume of information on the internet, information overload becomes a critical issue on many Web applications. Innumerable texts, audios, images and videos is faced by us every day. However, most of them are not intend to be received. Recommender systems play an indispensable role to help users alleviate the problem of information overload and boost user experience.

A session contains information that the user has inputted and which can be applied to track the movements of the user within the website without register or login. For instance, the information can be the timestamp and item id when a user clicks a certain item. Many people aren't willing to register or login due to inconvenient procedure and worry about privacy leakage. Conversely, most conventional recommender systems are content-based and collaborative filtering based ones. They manage to model users' preference to items on the basis of interactions between users and items. They utilized a user's historical interactions to learn the users' static long-term preference with the assumption that the importance of all historical interactions is equal. However, users' next action not only rely on long-term preference, but also depend on current interest which may be inferred from recent interactions. Besides, the conventional approaches ignore the sequential dependencies among the user's interactions, which lead to inaccurate modeling of the user's preference. In this case, session-based recommendation has become increasingly popular both in academic research and practical applications.

The task of session-based recommendation is to predict the next item that users would click from the current clicks sequence. Many kinds of proposals for session-based recommendation methods have been developed due to the highly practical value. Markov chain is a classic approach which assumes that the next action depends on the last or last few behaviors. Only considering a few of last behaviors makes the Markov chain based models [1] unable to utilize the dependencies of behaviors in long sequences and also might suffer from data sparsity issues. Rendle et al. [2] optimizes a pairwise ranking objective function via stochastic gradient descent, which ignores the temporal dependence among interactions and only considers some low-order interactions of latent factors. In recent years, many researches [3-7] apply Recurrent Neural Networks (RNNs) for session-based recommendation and achieved much better performance than conventional methods. Hidasi et al. [3] apply RNNs with ranking loss for session-based recommendation. Li et al. [4] propose NARM which models the user's sequential behavior and main purpose simultaneously. Liu et al. [5] propose STAMP which is capable of capturing users' general interests from the long-term memory of a session context. Ren et al. [6] take repeat consumption phenomenon into account and introduce repeat-explore mechanism, which significantly improve the performance under repeat session scenario. Wang et al. [7] bring collaborative modeling into this field with an end-to-end model. However, the RNNs based methods is hard to learn the dependencies from long distance. When the session sequence becomes longer, the performance of RNNs based methods decrease significantly. Recently, Graph neural networks (GNNs) have revolutionized the field of session-based recommendation which was dominated by recurrent neural networks (RNNs) by considering the transitions of items. Wu et al. [8] proposed SR-GNN which models session sequences as graph-structured data and capture complex transitions of items. Yu et al. [9] propose a target attentive network which discover the relevance of target item with graph neural network. Chen et al. [10] tackle information loss of GNN-based model by introducing shortcut graph attention and edge-order preserving aggregation layers. However, the GNNs based methods only learn the dependencies among adjacent items of the session graphs which is constructed from session sequences. When the items nodes are not adjacent in the session graph, their dependencies are hard to capture with GNNs even if their dependencies cannot be ignored.

Although the methods above achieved promising results and become the state-of-the-arts [11], the recurrent neural networks are hard to learn item dependencies from long distance. The graph neural networks based methods only aggregate the information from adjacent items where items are closely connected in the session graph. An attention based model have merge in the field of nature langue processing, named the Transformer proposed by Vaswani et al. [12]. It achieved excellent performance in the WMT 2014 [13] English-to-French and English-to-German translation tasks. The key factor of its success is self-attention mechanism which allow the model capture the dependencies between encoder and decoder. However, the encoder-decoder architecture of the Transformer is not suit for session-based recommendation which output size is not equal to input size. Xu et al. [14] apply self-attention layers after graph neural network to learn long-range dependencies among the output of graph neural network. However, the dependencies among all items may lost during the aggregation of adjacent items with the graph neural network. Adjacent items in a session graph may not be close in latent space, and vice versa. This may lead to inaccurate item embedding learning. The self-attention layer is directly applied after embedding layer to reserve the global item dependencies in SR-SAN.

The main contributions of this work are summarized as follows. Firstly, a self-attention based model which captures and reserves the full dependencies among all items regardless of their distance is proposed without using RNNs or GNNs. Secondly, to generate session-based recommendations, the proposed method use a single item latent vector which jointly represents current interest and global interest instead of session embedding which is composed of current interest embedding and global interest embedding as in [4,8-10,14]. In RNNs or GNNs based methods, the global interest embedding usually obtained by aggregating all items in the session with attention mechanism which is based on current interest embedding. However, this is redundant in SR-SAN which last item embedding is aggregating all items in the session with self-attention mechanism. In this way, the last item embedding in session can jointly represent current interest and global interest. Finally, extensive experiments have been conducted on real-world datasets without any assumptions. The experimental result shows that the developed method outperforms some state-of-the-arts by comparisons.

To make our results reproducible, the Pytorch version implementation is available at https://github.com/GalaxyCruiser/SR-SAN.

## 2. Session-based recommendation in SR-SAN

The proposed SR-SAN is introduced in this section. It utilizes the self-attention to learn global item dependencies. The multi-head attention mechanism is adopted to allow SR-SAN focus on different important part of the session. The latent vector of the last item in the session is used to jointly represents current interest and global interest with prediction layer.

The architecture of the proposed SR-SAN is shown in Figure 1.

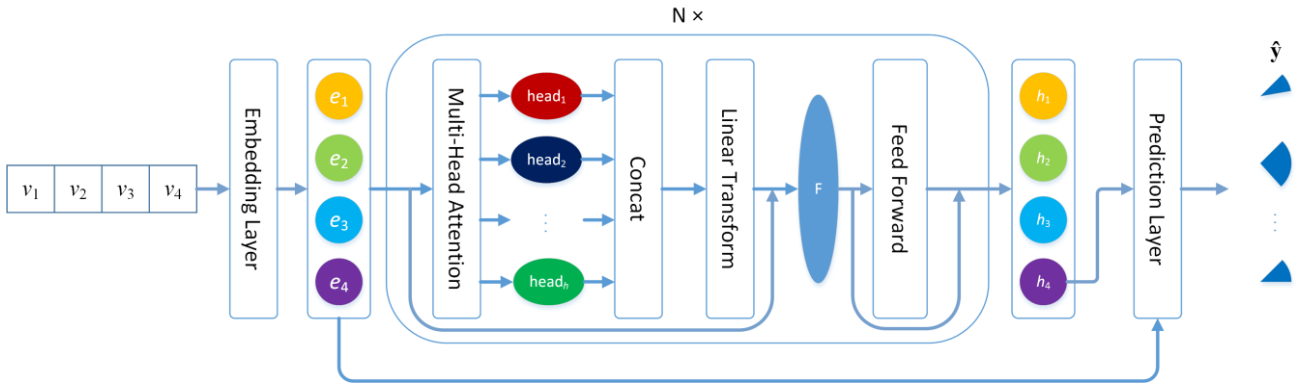

Figure 1: The architecture of SR-SAN

### 2.1 Problem Statement

Session-based recommender system makes prediction based upon current user sessions data without accessing to the long-term preference profile. Let $V = \{v_1, v_2, \ldots, v_{|V|}\}$ denote the set consisting of all unique items involved in all the sessions. An anonymous session sequence $S$ can be represented by a list $S = [s_1, s_2, \ldots, s_n]$, where $s_i \in V$ represents a clicked item of the user within the session $S$. The task of session-based recommendation is to predict the next click $s_{n+1}$ for session $S$. Our models are constructed and trained as a classifier that learns to generate a score for each of the candidates in $V$. Let $\hat{\mathbf{y}} = \{\hat{y}_1, \hat{y}_2, \ldots, \hat{y}_{|V|}\}$ denote the output score vector, where $\hat{y}_i$ corresponds to the score of item $v_i$. The items with top-$K$ values in $\hat{\mathbf{y}}$ will be the candidate items for recommendation.

### 2.2 Model Overview

The proposed model is made up of two parts. The first part is obtaining item latent vectors with self-attention networks, the second part of the proposed model is making recommendation with prediction layer.

### 2.3 Learning Item Embeddings with Self-Attention Network

First, the input session $S = [s_1, s_2, \ldots, s_n]$ is passed through an embedding layer to obtain the corresponding $d$ dimensional embeddings $\mathbf{E} = [\mathbf{e}_1, \mathbf{e}_2, \ldots, \mathbf{e}_n]$. After that, item embeddings are fed into the self-attention network. The self-attention network is made up of multi-head attention networks and feedforward network.

An attention function can be described as mapping a query vector and a set of key-value vector pairs to an output vector. The output is calculated as a weighted sum of values, where the weight assigned to each value is calculated by querying the compatibility function with the corresponding key. The self-attention network adopted the multi-head attention mechanism which allows different attention heads attend to information at different positions.

The $i$th head attention is computed as:

$$\text{head}_i = \text{softmax}(\frac{(\mathbf{EW}_i^Q)(\mathbf{EW}_i^K)^T}{\sqrt{d}})(\mathbf{EW}_i^V) \quad (1)$$

Where the $\mathbf{W}_i^Q, \mathbf{W}_i^K, \mathbf{W}_i^V \in \mathbb{R}^{d \times (d/h)}$ are parameter matrices of the $i$th head attention, $h$ is the number of attention heads. The dot products of query and key is divided by $d^{1/2}$, pushing the softmax into regions where it has bigger gradients than not scaled that would alleviate gradient vanish. Note that, the same $\mathbf{E}$ is used in the projections of query, key and value matrices, which is called self-attention. Each item in the session is adaptively assign weights to all other items with attention mechanism. In this way, each item can learn the dependencies with all other items. Meanwhile, such parallel matrix multiplication method is more efficient than the recurrent manner.

Then, the $h$ heads of self-attention are combined with multi-head attention. Multi-head attention allows the model to jointly learn the dependencies at different positions. The residual connection is also adopted in the attention network to help the model learn low-layer information. The multi-head attention $\mathbf{F}$ is computed as:

$$\mathbf{F} = \text{Concat}(\text{head}_1, \cdots, \text{head}_h)\mathbf{W}^O + \mathbf{E} \quad (2)$$

Where $\mathbf{W}^O \in \mathbb{R}^{d \times d}$ is the projection matrix.

Then, the $\mathbf{F}$ is passed through a two layer fully connected feed-forward network with an activation function in the middle. Residual connection is also adopted in feedforward network.

$$\mathbf{H} = \max(0, \mathbf{FW}_1 + \mathbf{b}_1)\mathbf{W}_2 + \mathbf{b}_2 + \mathbf{F} \quad (3)$$

Where $\mathbf{H} = [\mathbf{h}_1, \mathbf{h}_2, \ldots, \mathbf{h}_n]$ is the latent vectors of the items in the session.

**2.4 Making Recommendation**

The prediction layer is composed of score compute with dot product and softmax function. The score $\hat{z}_i$ for each candidate item $v_i \in V$ is computed after we obtained the latent vector of last item in session $S$ as:

$$\hat{z}_i = \mathbf{h}_n^T \mathbf{e}_i \quad (4)$$

$\mathbf{h}_n$ is the last item's latent vector which jointly capture the user's current interest and global interest. Because the last item's adaptively aggregation from all other items with attention mechanism.

The output vector of model $\hat{\mathbf{y}}$ is obtained by apply a softmax function on $\hat{\mathbf{z}}$ as:
$$\hat{\mathbf{y}} = \text{softmax}(\hat{\mathbf{z}}) \qquad (5)$$
where $\hat{\mathbf{z}} \in \mathbb{R}^{|V|}$ denotes the recommendation scores of all candidate items. $\hat{\mathbf{y}}$ denotes the probability of items to be next clicked.

### 2.5 Training

The model is trained by minimizes the cross-entropy between the ground-truth and the prediction:
$$\mathcal{L}(\hat{\mathbf{y}}) = -\sum_{i=1}^{|V|} \mathbf{y}_i \log(\hat{\mathbf{y}}_i) + (1-\mathbf{y}_i)\log(1-\hat{\mathbf{y}}_i) \qquad (6)$$
where $\mathbf{y}$ denotes the one-hot encoding vector of the ground truth item. This function is optimized with the back-propagation through time (BPTT) algorithm.

## 3. Experiments and Discusses

Some datasets, baseline methods and evaluation metrics used in the experiments are described firstly in this section. The proposed method has been tested with some state-of-the-arts at the selected datasets.

### 3.1 Datasets

Two representative datasets including Yoochoose [15] and Diginetica [16] are selected to test the performance of the proposed method. Yoochoose [15] is a challenge dataset of RecSys Challenge 2015 which contains a record of click-streams from an E-commerce website within 6 months. Diginetica [16] is another challenge dataset of CIKM cup 2016, only transaction data in used in the experiment.

For a fair comparison, sessions of length 1 and items appearing less than 5 times in both datasets are filtered out as [3, 4, 5]. The Yoochoose [15] remains 7,981,580 sessions and 37,483 items, while the Diginetica [16] dataset remains 204,771 sessions and 43,097 items. Data augment is adopted by split a session of length $n$ into $n$-1 sessions of length ranging from 2 to $n$ as [17]. The last item is the label item. For the Yoochoose [15], the most recent 1/64 portions of the training sequences are used.

Table 1: Statics of the experiment datasets

| Dataset | All the clicks | Train sessions | Test sessions | All the items | Avg. length |
|---|---|---|---|---|---|
| Yoochoose1/64 [15] | 557,248 | 369,859 | 55,898 | 16,766 | 6.16 |
| Diginetica [16] | 982,961 | 719,470 | 60,858 | 43,097 | 5.12 |

### 3.2 Baselines

The proposed method is compared with the following representative ones to evaluate the performance. Item-KNN [18] recommends similar items of the previous clicked item in the session based on cosine similarity. BPR-MF [2] optimizes a pairwise ranking objective function via stochastic gradient descent. FPMC [1] models next-basket recommendation. In order to adapt to session-based recommendation, we omit the unavailable user feature. GRU4REC [3] models user sequences for session-based recommendation using RNNs. NARM [4] employs a local encoder and a global encoder with an attention mechanism to model the user's sequential behavior and capture the user's main purpose. STAMP [5] captures users' general interests of session context and users' current interests of last click. Ren et al. [6] take repeat consumption phenomenon into account. Wang et al. [7] bring collaborative modeling into this field with an end-to-end model. SR-GNN [8] models session sequences into graph-structure data and uses graph neural networks to capture complex item transitions. Yu et al. [9] propose a target attentive network which discover the relevance of target item with graph neural network. Chen et al. [10] tackle information loss of GNN-based model by introducing shortcut graph attention and edge-order preserving aggregation layers. Xu et al. [14] apply self-attention layers after graph neural network to learn long-range dependencies.

### 3.3 Evaluation Metrics

Since recommender system can only recommend a few items at one time, the following metric is adopted for evaluation of the session-based one, which is widely used in related works.

HR@20 (Hit Rate) represents the proportion of correctly recommended items amongst the top-20 items in all test cases.

$$\text{HR@}20 = \frac{n_{\text{hit}}}{N}, \tag{7}$$

where $N$ represents the number of test sequence in the dataset and $n_{\text{hit}}$ represents the number that the desired items are at the top 20 position in the ranking list.

MRR@20 (Mean Reciprocal Rank) represents the average of reciprocal ranks of the desired items. The reciprocal rank is set to 0 when it exceeds 20.

$$\text{MRR@}20 = \frac{1}{N} \sum_{v_{\text{label}} \in \mathbb{S}_{\text{test}}} \frac{1}{\text{Rank}(v_{\text{label}})} (\text{Rank}(v_{\text{label}}) \leq 20), \tag{8}$$

where $v_{\text{label}}$ is the label of the session, $\mathbb{S}_{\text{test}}$ is the labels of test set.

### 3.4 Experiments Setting

A Gaussian distribution with a mean of 0 and a stand deviation of 0.1 is adopted to initialized all parameters of the SR-SAN. The Adam optimizer is adopted with the initial learning rate $1e-3$ and decay by 0.1 every 3 epochs. The batch size is set to 100. The L2 regularization is set to $1e-5$ to mitigate overfitting.

### 3.5 Comparisons with Some Baseline Methods

To demonstrate the performance of the proposed method, some state-of-the-arts as mentioned above are selected to test both in HR@20 and MRR@20, respectively. The overall performance is presented in Table 2, in which the best results are highlighted in boldface.

Table 2: Performance comparisons with some state-of-the-arts

| Methods | Yoochoose 1/64 [15] | | Diginetica [16] | |
|---|---|---|---|---|
| | HR@20 | MRR@20 | HR@20 | MRR@20 |
| Item-KNN [18] | 51.60 | 21.81 | 35.75 | 11.57 |
| BPR-MF [2] | 31.31 | 12.08 | 5.24 | 1.98 |
| FPMC [1] | 45.62 | 15.01 | 26.53 | 6.95 |
| GRU4REC [3] | 60.64 | 22.89 | 29.45 | 8.33 |
| NARM [4] | 68.32 | 28.63 | 49.70 | 16.17 |
| STAMP [5] | 68.74 | 29.67 | 45.64 | 14.32 |
| RepeatNet [6] | 70.71 | 31.03 | 47.79 | 17.66 |
| CSRM [7] | 71.45 | 30.36 | 50.55 | 16.38 |
| SR-GNN [8] | 70.57 | 30.94 | 50.73 | 17.59 |
| GC-SAN [14] | 70.66 | 30.04 | 51.70 | 17.61 |
| TAGNN [9] | 71.02 | 31.12 | 51.31 | 18.03 |
| LESSR [10] | 70.64 | 30.97 | 51.71 | **18.15** |
| **Ours** | **71.74** | **31.58** | **52.04** | 17.61 |

One can find that the proposed method overall performance outperforms the investigated state-of-the-arts by comparisons from Table 2. Some reasons are discussed as follows. BPR-MF [2] only models the low-order interactions of latent factors and ignores the sequence dependencies. The performance of FPMC [1] is much better than that of BPR-MF [2], the performance, however, is still undesirable. It may attribute to the assumption on the independence of successive items, which is not realistic. Since Item-KNN [18] utilizes the similarity between items, it outperforms both FPMC [1] and BPR-MF [2]. It is undesirable also due to the sequential information being ignored in Item-KNN [18]. GRU4REC [3] applied RNNs at first, it outperforms the methods [1, 2, 18] as a whole. Both NARM [4] and STAMP [5] consider the current interests and general interests, they get the better performance against some ones [1, 2, 3, 18] as mentioned above. RepeatNet [6] take repeat consumption phenomenon into account, improve the performance under repeat session scenario, is better than previous RNNs based methods. CSRM [7] apply collaborative neighbor information to current session, achieved the second place on Yoochoose dataset in term of HR@20, the performance is still comparable on other metrics with other deep learning based methods. However, RNNs based methods are hard to learn long-range dependencies. SR-GNN [8] models session sequences as graphs and further consider the transitions of items, the overall performance is better than RNN-based methods. TAGNN [9] is developed from the SR-GNN [8] with target attentive module, it outperforms the SR-GNN [8] on all metrics. LESSR [10] is the best on the Diginetica dataset in term of MRR@20, and HR@20 is second place in the investigated methods with the handling of information loss of GNN-based methods by introducing shortcut graph attention and edge-order preserving aggregation layers. LESSR [10] do not provide the result on the Yoochoose dataset. The experiment is conducted using the code they provide with default settings. The result on the Yoochoose is not outstanding as it on the Diginetica dataset, but still much better than traditional methods. GC-SAN [14] apply self-attention layers after graph neural network to capture long-range dependencies. However, the global item dependencies are lost after the neighbor item aggregation with graph neural network. But GC-SAN [14] is still outperform the SR-GNN [8] with the self-attention layer. The GNN-based models only consider the relation between adjacent items, neglect the relation between remote items whether they are close in latent space or not that restrict the capability of model the dependencies of items. The proposed method utilizes the self-attention mechanism to learn the dependencies among all items other than the dependencies of adjacent items. Such mechanism lead to better item embedding learning. Moreover, the latent vector of last item in the session in used to jointly represent the current interest and global interest in the prediction layer. The SR-SAN achieve the best performance on the Yoochoose dataset in term of HR@20 and MRR@20 and on the Diginetica dataset in term of HR@20. The performance on the Diginetica dataset in term of MRR@20 is still comparable with other state-of-the-art methods.

**3.6 Comparison with Session Embedding Variant**

Previous methods usually capture current interest and global interest with the composition of current interest embedding and global interest embedding in prediction layer. The SR-SAN is compared with the same session embedding strategy (SR-SAN-SE) used in the previous methods.

From Table 3, it can be observed that the session embedding variant is inferior to SR-SAN, which validates the redundancy of the composition of current interest embedding and global interest embedding in self-attention networks. The global interest embedding is obtained by aggregating all items in the session with attention mechanism based on the current interest embedding. The last item's latent vector in self-attention network is obtained in the similar way. Thus, the last item's latent vector can jointly represent current interest and global interest. The redundancy of global interest embedding leads to performance downgrade.

Table 3: Performance of different prediction schemes

| Methods | Yoochoose 1/64 [15] | | Diginetica [16] | |
| --- | --- | --- | --- | --- |
|  | HR@20 | MRR@20 | HR@20 | MRR@20 |
| SR-SAN-SE | 71.07 | 31.34 | 51.14 | 17.23 |
| **SR-SAN** | **71.74** | **31.58** | **52.04** | **17.61** |

### 3.7 Comparison with GNNs based Variant

The proposed method is compared with the variant which apply the graph neural network that SR-GNN [8] adopt before the self-attention network (SR-SAN-GNN). From Table 4, one can notice that, the involve of graph neural network make the performance of the model downgrade. The graph neural network focus on the adjacent items that may lose the dependencies among remote items.

Table 4: Performance of different architectures

| Methods | Yoochoose 1/64 [15] | | Diginetica [16] | |
| --- | --- | --- | --- | --- |
|  | HR@20 | MRR@20 | HR@20 | MRR@20 |
| SR-SAN-GNN | 71.34 | 31.50 | 51.82 | 17.32 |
| **SR-SAN** | **71.74** | **31.58** | **52.04** | **17.61** |

### 3.8 Hyper-parameter Study

In this section, how the embedding size, heads of attention, dimension of feedforward neural network and the layers of SAN influence the proposed model is investigated. Due to space limit, only result in term of HR@20 is shown.

The result of how the embedding size influent the performance is shown in Figure 2. From the figure, one can notice that increasing the size of embedding not always improve the performance of the proposed model. For the Diginetica dataset, the optimal embedding size is 48, of which data size is smaller than the Yoochoose dataset. For the Yoochoose dataset, the optimal embedding size is 96. The performance deteriorates quickly when the embedding size is bigger than the optimal size. The reason is the too big size of embedding cause the model overfitting. The performance is also down grade when the embedding is smaller than the optimal embedding size due to limited learning capacity.

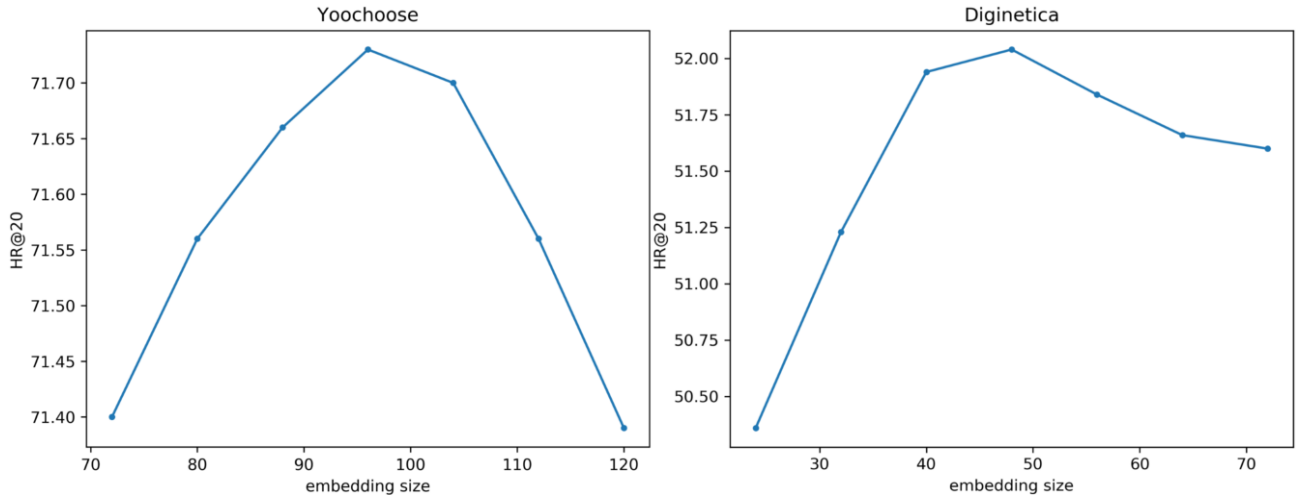

Figure 2: The performance under different embedding size.

In Figure 3, the influence of attention heads is investigated. Note that the embedding size is a multiple of attention heads. So the number of attention heads is not consecutive. Multi-head attention allow the model attend to different positions of the sequence. However, too large attention heads make the single attention dimension too small that limit the expressive power of attention. The optimal attention heads of Yoochoose is 2 and the optimal attention heads of Diginetica is 8. The performance is slightly reduced when the attention heads larger or smaller than the optimal value.

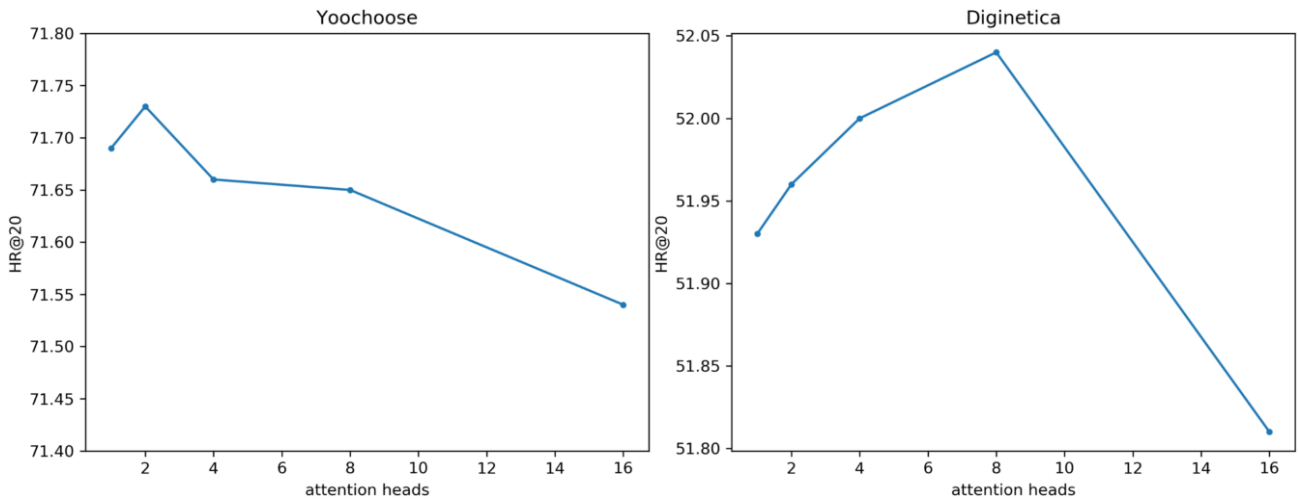

Figure 3: The performance under different attention heads.

In Figure 4, the number of SAN layers is investigated. From the figure, one can notice that the performance down grade with the increase of SAN layers. The optimal number of SAN layers is 1. SR-SAN may suffer from overfitting with larger number of SAN layers.

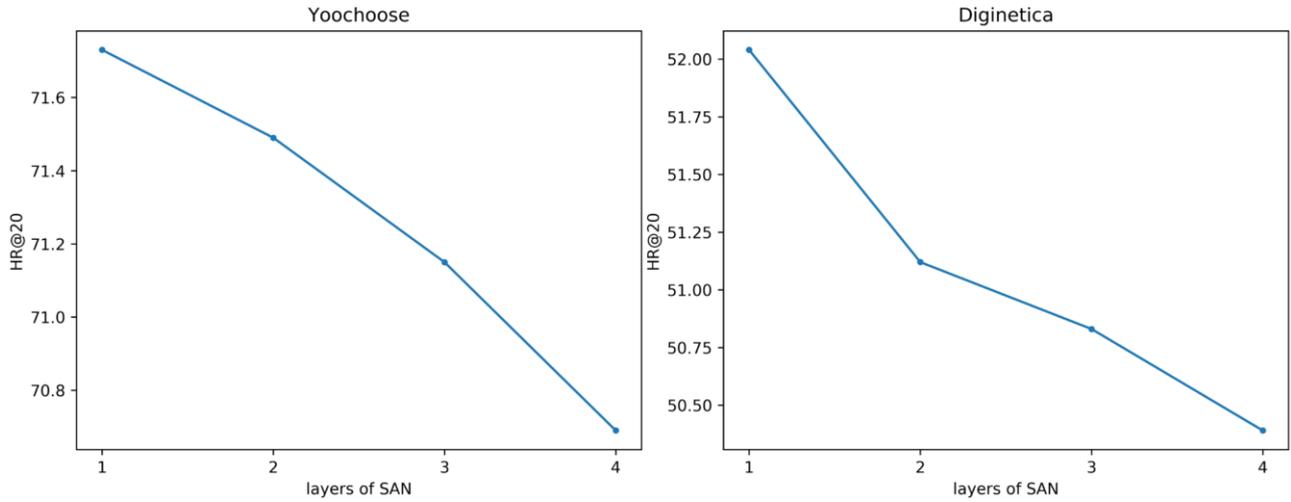

Figure 4: The performance under different layers of SAN.

In Figure 5, the dimension of feedforward neural network is investigated. The x axis is the multiplier of embedding size. One can notice that the optimal multiplier is 4. SR-SAN suffer from underfit when the multiplier is smaller than 4 and SR-SAN suffer from overfit when the multiplier larger than 4.

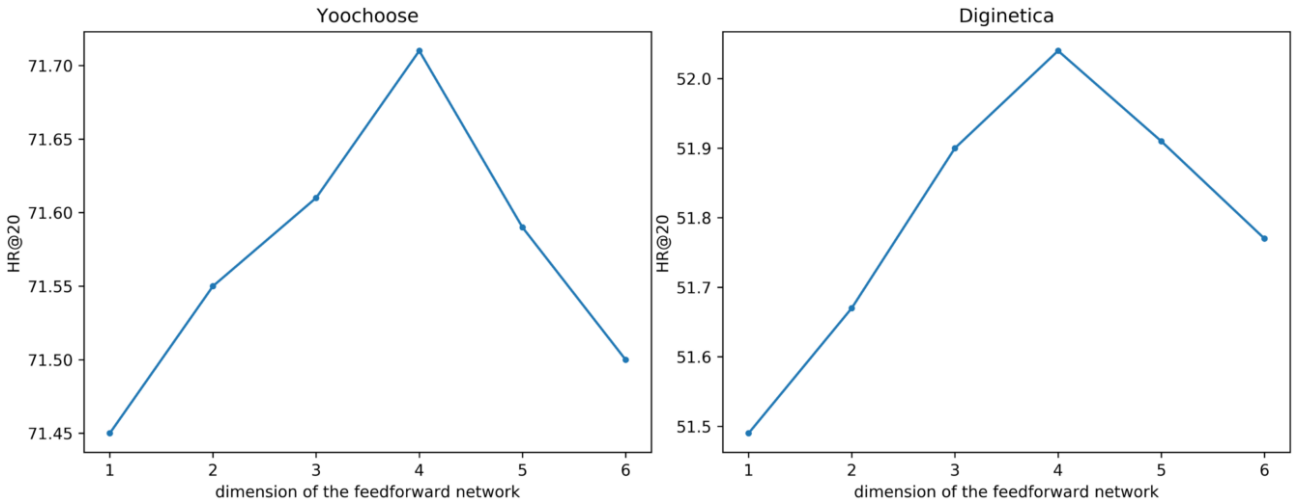

Figure 5: The performance under different dimension of feedforward network.

## 4. Conclusions

A novel session-based recommendation has been proposed to generate recommendation results from anonymous sessions. The developed SR-SAN utilizes the self-attention network to learn the global item dependencies. The proposed SR-SAN is not only able to learn the dependencies among adjacent items, but also able to learn the dependencies among remote items regardless of their distance. The latent vector of last item is used to jointly represents the current interest and global interest instead of session embedding which composed of current interest embedding and global interest embedding. The proposed SR-SAN is a basic implement of self-attention network based method for session-based recommendation. Only the latent vector corresponding to the last item of the session is used in prediction layer. Some experiments have been done to test the performance of the developed method at real-world datasets without any assumptions. The experimental result shows that the developed method outperforms some state-of-the-arts by comparisons.

**Reference**

[1] S. Rendle, C. Freudenthaler and L. Schmidt-Thieme. Factorizing personalized Markov chains for next-basket recommendation. Proceeding of International World Wide Web Conference, ACM, 2010: 811-820.

[2] S. Rendle, C. Freudenthaler, Z. Gantner, et al. BPR: Bayesian personalized ranking from implicit feedback. Proceeding of uncertainty in artificial intelligence, 2009: 452-461.

[3] H. Balázs, M. Quadrana, A. Karatzoglou, et al. Parallel recurrent neural network architectures for feature-rich session-based recommendations. Proceedings of the 6th ACM Conference on Recommender Systems, ACM, 2016: 241-248.

[4] J. Li, P. Ren, Z. Chen, et al. Neural attentive session-based recommendation. Proceedings of the 2017 ACM on Conference on Information and Knowledge Management, ACM, 2017: 1419-1428.

[5] Q. Liu, Y. Zeng, R. Mokhosi, et al. Stamp: short-term attention/memory priority model for session-based recommendation. Proceedings of the 24th ACM SIGKDD International Conference on Knowledge Discovery and Data Mining, ACM, 2018: 1831-1839.

[6] P. Ren, Z. Chen, J. Li, et al. RepeatNet: A repeat aware neural recommendation machine for session-based recommendation. In Proceedings of the AAAI Conference on Artificial Intelligence 2019: 4806-4813.

[7] M. Wang, P. Ren, L. Mei, et al. A Collaborative Session-based Recommendation Approach with Parallel Memory Modules. In Proceedings of the 42nd International ACM SIGIR Conference on Research and Development in Information Retrieval, 2019: 345–354.

[8] S. Wu, Y. Tang, Y. Zhu, et al. Session-based recommendation with graph neural networks. Proceedings of the 33rd AAAI Conference on Artificial Intelligence, 2019: 346-353.

[9] F. Yu, Y. Zhu, Q. Liu, et al. TAGNN: Target Attentive Graph Neural Networks for Session-based Recommendation. In Proceedings of the 43rd International ACM SIGIR Conference on Research and Development in Information Retrieval, 2020:1921-1924.

[10] T. Chen and R. C. Wong. Handling Information Loss of Graph Neural Networks for Session-based Recommendation. In Proceedings of the 26th ACM SIGKDD International Conference on Knowledge Discovery & Data Mining, 2020: 1172–1180.

[11] H. Fang, G. Guo, D. Zhang, et al. Deep learning-based sequential recommender systems: concepts, algorithms, and evaluations. Proceeding of International Conference on Web Engineering, 2019: 574-577.

[12] A. Vaswani, N. Shazeer, N. Parmar, et al. Attention is all you need. In Advances in neural information processing systems, 2017: 5998-6008.

[13] http://www.statmt.org/wmt14/

[14] C. Xu, P. Zhao, Y. Liu, et al. Graph Contextualized Self-Attention Network for Session-based Recommendation. Proceedings of the 28th International Joint Conference on Artificial Intelligence, 2019: 3940-3946.

[15] http://2015.recsyschallenge.com/challege.html

[16] http://cikm2016.cs.iupui.edu/cikm-cup

[17] Y. K. Tan, X. Xu, and Y. Liu. Improved recurrent neural networks for session-based recommendations. Proceedings of the 1st Workshop on Deep Learning for Recommender Systems, 2016: 17-22.

[18] B. Sarwar, G. Karypis, J. Konstan, et al. Item-based collaborative filtering recommendation algorithms. Proceeding of International World Wide Web Conference, 2001: 285-295.